\documentclass[a4paper]{article}
\usepackage{amssymb}
\usepackage{natbib}
\usepackage{hyperref}
\usepackage{graphicx}

\def \O{{\rm O}}
\def\Z{Z}
\def\Zs{{\check Z}}
\def\Ss{{\check S}}
\def \1{{\rm Id}}
\def \2{{\cal I}}
\def \rank{{\rm rank\,}}
\def \trace{{\rm \,trace\,}}
\def\d{{\rm d}}

\newcommand{\C}{\mathbb{C}}
\newcommand{\N}{\mathbb{N}}
\newcommand{\R}{\mathbb{R}}
\newcommand{\T}{\mathbb{T}}

\title{\large{\bf A limit of nonplanar 5-body central configurations is nonplanar}}

\author{}

\date{}

\begin{document}
	
	\maketitle

	\begin{center}
	{\bf Alain Albouy$^{1}$,\qquad Antonio Carlos Fernandes$^{2}$}
	
	\bigskip
	$^{1}$ IMCCE, CNRS-UMR8028, Observatoire de Paris
	
        77, avenue Denfert-Rochereau, 75014 Paris, France
         
         Alain.Albouy@obspm.fr
         
         \bigskip
         
         $^{2}$ Instituto de Matem\'atica e Computa\c c\~ao,
         
         Universidade Federal de Itajub\'a, Minas Gerais, Brazil 
         
acfernandes@unifei.edu.br
        
        \bigskip
	\end{center}

{\bf Abstract.}  \citet{moeckel}, \citet{moeckelsimo} proved that, while continuously changing the masses, a 946-body planar central configuration bifurcates into a spatial central configuration. We show that this kind of bifurcation does not occur with 5 bodies.
Question 17 in the list \citet{ACS} is thus answered negatively.

\bigskip

\centerline{\bf 1. Introduction}

\bigskip

The aim of this article is to prove the following Theorem, where $n$-body configurations are thought of as configurations of positions $(q_1,\dots,q_n)\in(\R^p)^n$ together with positive masses $(m_1,\dots,m_n)\in\R^n$. The positive integer $p$ is the dimension of the position space. The dimension $d$ of a configuration is the dimension of the space generated by the vectors $(q_2-q_1,q_3-q_1,\dots,q_n-q_1)$. We have $d\leq p$ and $d\leq n-1$. By ``limit'' we mean the limit of a converging sequence in the space $(\R^p)^n\times \R^n$. Each position tends to the limiting position and each mass tends to a limiting mass, which is assumed to be positive.

{\bf 1.1. Theorem.} A 5-body central configuration which is the limit of a sequence of 5-body central configurations of dimension $d$ is  of dimension $d$.

The corresponding result is true for $n\leq 4$, unknown for $6\leq n\leq 945$, false for $n=946$ (see \S 4.3). Recall that a configuration of $n$ bodies with given positive masses is {\it central} if and only if it is, for some $\lambda>0$, a critical point of the amended Newtonian force function $U+\lambda I/2$, where $$U=\sum_{1\leq i<j\leq n}\frac{m_im_j}{r_{ij}},\quad I=\frac{1}{M}\sum_{1\leq i<j\leq n}m_im_jr_{ij}^2,\eqno(1.1)$$
$$M=m_1+\cdots+m_n,\quad r_{ij}=\|q_i-q_j\|.$$
The configuration considered in the next theorem is a planar central configuration with $n=5$ and $\lambda=M$. The choice $\lambda=M$ does not restrict the generality. It simply scales the central configurations, in such a way that for example the equilateral Lagrange configuration ($n=3$, $d=2$) has its three sides equal to 1, whatever the masses.

{\bf 1.2. Theorem.} Consider in $\R^3$ a 5-body planar configuration with positions $q_i\in\R^3$ and masses $m_i>0$, $i=1,\dots,5$, which is a critical point of the function $\sum_{i<j}m_im_j(r_{ij}^{-1}+r_{ij}^2/2)$, where $r_{ij}=\|q_i-q_j\|$. Let an orthonormal frame $\O xyz$ be such that $q_i=(x_i,y_i,0)$. A $(v_1,\dots,v_5)\in(\R^3)^5$, with $v_i=(0,0,\zeta_i)$, is in the kernel of the Hessian quadratic form of this function if and only if $(\zeta_1,\dots,\zeta_5)$ is a linear combination of $(1,\dots, 1)$, $(x_1,\dots,x_5)$ and $(y_1,\dots,y_5)$.

Note that these linear combinations generate infinitesimal translations and rotations of the configuration. Let us compare our result with the following one.

{\bf 1.3. Theorem (\citealt{moeckel}).} Every planar central configuration of $n\geq 4$ masses has at least one negative eigenvalue with eigenvector normal to the manifold of planar configurations.

Let us call a $(v_1,\dots,v_n)$ with all the vectors $v_i$ orthogonal to the configuration a {\it vertical vector}. Moeckel proved that for an arbitrary planar $n$-body central configuration, $n\geq 4$, there is a vertical eigenvector of the Hessian quadratic form with a negative eigenvalue. We state in 1.2 that if $n=5$ two independent vertical eigenvectors have a nonzero eigenvalue\footnote{We believe that both eigenvalues are always negative. This should be checked at one point of each connected component of the subvariety of $(\R^2)^5\times(\R_{>0})^5$  formed by the central configurations. But we do not know the number of connected components. The cited result of \citet{chh} and \citet{gidea} is relevant for this question. The cited result of \citet{asv} gives the negative definiteness in a particular case.}. There cannot be more than two since the maximal number of such independent eigenvectors is $n-3$. Theorem 1.2 together with the following proposition gives Theorem 1.1.

{\bf 1.4. Proposition.} If an $n$-body central configuration of dimension $d$ is the limit of a sequence of $n$-body central configurations of dimension $>d$, then the Hessian quadratic form of the amended force function has a nontrivial vertical degeneracy. 

By a vertical degeneracy we mean a vertical vector in the kernel of the Hessian quadratic form. By nontrivial we mean that this vector does not generate a translation or a rotation. The next results are elementary consequences of a block diagonal structure of the Hessian matrix.

{\bf 1.5. Proposition.} If $(u_1,\dots,u_n)\in(\R^p)^n$ belongs to the kernel of the Hessian quadratic form of the amended force function, and if we project each $u_i$ on a vertical direction, we obtain a vertical vector in this kernel.

{\bf 1.6. Proposition.} A vertical vector $(u_1,\dots,u_n)\in(\R^p)^n$ belongs to the kernel of the Hessian quadratic form of the amended force function if and only if the list of its components $(z_1,\dots,z_n)\in\R^n$ on any vertical direction belongs to the kernel of the shifted Wintner-Conley matrix.

We will prove these results in reverse order in the next sections. We will first recall the construction of the Wintner-Conley matrix and of the shifted Wintner-Conley matrix. We will show in a future work that a nontrivial degeneracy of the latter has surprising consequences if $d=n-3$.

\bigskip
\centerline{\bf 2. The Wintner-Conley matrix.}
\bigskip

Here we prove Propositions 1.5 and 1.6 after recalling some known facts. The Newton equations of the $n$-body problem are $$\ddot q_i=-\gamma_i, \qquad i=1,\dots,n,$$ where
$$\gamma_i=\sum_{j\neq i}m_jS_{ij}(q_i-q_j),\qquad S_{ij}=r_{ij}^{-3},\qquad r_{ij}=\|q_i-q_j\|.$$

{\bf 2.1. Definition.} A configuration of point bodies with positive masses is called a {\it central configuration} if there exist $\lambda\in\R$ (the multiplier) and $q_0\in\R^p$ (the center of attraction) such that $\lambda(q_i-q_0)=\gamma_i$.

{\bf 2.2. Proposition.} In a central configuration with positive masses, the center of attraction $q_0$ is the center of mass of the configuration, and the multiplier $\lambda$ is positive. 

{\bf Proof.} We have $\sum m_i\gamma_i=0$ and consequently $\lambda(q_G-q_0)=0$ where $q_G=(\sum m_i q_i)/M$ is the center of mass. Consider a body $q_i$ on the boundary of the convex hull of the configuration. The attraction vector $-\gamma_i$ is the sum of vectors all pointing toward the same half-space, which contains the center of mass. So $\lambda\neq 0$,  $q_0=q_G$ and finally $\lambda>0$. QED

{\bf 2.3. Proposition.} A configuration with positive masses is central with multiplier $\lambda$ if and only if it is a critical point of $U+\lambda I/2$.

{\bf Proof.} From the formulas $(1.1)$ we compute the derivative of $U+\lambda I/2$ in the direction $u=(u_1,\dots,u_n)$:  $$\langle \d(U+\lambda I/2),(u_1,\dots,u_n)\rangle=-\sum_{i,j,i<j}m_im_j(S_{ij}-{\lambda\over M})\langle q_i-q_j,u_i-u_j\rangle\eqno(2.1)$$
$$=-\sum_{i,j,i\neq j}m_im_j(S_{ij}-{\lambda\over M})\langle q_i-q_j,u_i\rangle=-\sum_{i=1}^n m_i\langle \gamma_i-\lambda(q_i-q_G),u_i\rangle.$$
This quantity should be zero for all $u$. QED

{\bf Hessian quadratic form.} Let us differentiate once more and express the Hessian bilinear form
$$\langle \d^2(U+\lambda I/2),(u_1,\dots,u_n)\otimes(v_1,\dots,v_n)\rangle=-\sum_{i,j,i<j}m_im_j(S_{ij}-{\lambda\over M})\langle u_i-u_j,v_i-v_j\rangle$$
$$+3\sum_{i,j,i<j}m_im_jr_{ij}^{-5}\langle q_i-q_j,u_i-u_j\rangle \langle q_i-q_j,v_i-v_j\rangle$$
$$=\sum_{i,j,i\neq j}m_im_j\bigl(-(S_{ij}-{\lambda\over M})\langle u_i-u_j,v_i\rangle+3r_{ij}^{-5}\langle q_i-q_j,u_i-u_j\rangle \langle q_i-q_j,v_i\rangle\bigr)$$
$$=\sum_i m_i\langle g_i,v_i\rangle\quad\hbox{with}$$
$$g_i=\sum_{j,j\neq i}m_j\bigl(-(S_{ij}-{\lambda\over M})(u_i-u_j)+3r_{ij}^{-5}\langle q_i-q_j,u_i-u_j\rangle(q_i-q_j)\bigr).\eqno(2.2)$$
The configuration is conventionally horizontal. If we take horizontal $u_i$'s and vertical $v_i$'s the bilinear form is zero. If both the $u_i$'s and the $v_i$'s are vertical, there is only the first term, where there are only first derivatives of $U$. This term may be expressed with the {\it Wintner-Conley matrix.}

The Wintner-Conley matrix $\Z$ gives a matrix form to the Newton equations $\ddot q_i=-\gamma_i$. It gives the $\gamma_i$'s as follows
$$(\gamma_1,\dots,\gamma_n)=(q_1,\dots,q_n)\Z$$
where each $q_i$ and each $\gamma_i$ is a column vector and $\Z$ is the matrix
$$\Z=\pmatrix{\Sigma_1& -m_1 S_{12}&-m_1 S_{13}&\dots& -m_1S_{1n}\cr -m_2S_{12}&\Sigma_2&-m_2S_{23}&\dots &-m_2S_{2n}\cr -m_3 S_{13}& -m_3 S_{23}&\Sigma_3&\dots& -m_3S_{3n}\cr \dots&\dots&\dots&\dots&\dots
\cr
-m_n S_{1n}&-m_n S_{2n}&-m_n S_{3n}&\dots&\Sigma_n},\qquad \Sigma_i=\sum_{j\neq i}m_jS_{ij}.\eqno(2.3)$$
The equation of motion is now $\ddot q=-q\Z$ while the equation of central configurations is $\lambda (q-q_G)=q \Z$, with, for a configuration in $\R^3=\O xyz$, 
$$q=(q_1,q_2,\dots, q_n)=\pmatrix{x_1&x_2&\cdots&x_n\cr
y_1&y_2&\cdots&y_n\cr z_1&z_2&\cdots&z_n}.\eqno(2.4)$$ We may remove the $q_G$ term by simply restricting to configurations with the center of mass at zero. Then the equation is $\lambda q=q \Z$. The rows of the matrix $q$ are eigenvectors with eigenvalue $\lambda$ of the matrix $\Z$ (conventionally acting at the right). If we replace $\Z$ by $\Z-\lambda \1$, where $\1$ is the identity matrix, the rows of $q$ are now in the left kernel. But $(1,1,\dots,1)$ is not in this kernel. We will prefer to define a {\it shifted Wintner-Conley matrix} with $(1,1,\dots,1)$ in its left kernel. The shifted matrix will then give the terms in $\lambda$ of $(2.1)$ and $(2.2)$.

{\bf 2.4. Quotient space.} A row of the matrix $q$ representing a configuration with center of mass at the origin is called a disposition. It is an element of the $n-1$ dimensional vector space of the $(x_1,\dots,x_n)\in \R^n$ such that $\sum m_ix_i=0$. Another definition was emphasized in \citet{albouychenciner} and \citet{albouy3} as giving more general and simpler equations. A disposition is a configuration $(x_1,\dots,x_n)$ on the real line, considered {\it up to translations}. The space of dispositions is now a quotient space instead of a subspace. A linear form on this space is an invariant linear form on $\R^n$, i.e., a $(\xi_1,\dots,\xi_n)\in\R^n$ such that $\sum_{i=1}^n\xi_i=0$. We see that the columns of $\Z$ satisfy this condition. If we take the equation $\lambda q=q\Z$ as the definition of a central configuration, and if we adopt the second definition of a disposition, then the new definition is correct for real $m_i$'s such $M=m_1+\cdots+m_n=0$. In the case, there is no center of mass. For example, an equilateral triangle is a central configuration for any $(m_1,m_2,m_3)\in\R^3$, even if $M=0$.

Instead of subtracting $\lambda\1$ to $\Z$, as Wintner indeed proposed, we introduce another matrix which induces the identity on the disposition space, namely, the matrix $\2$ such that $0=(1,\dots,1)\,\2$ and such that $x=x\,\2$ for any
$x=(x_1,\dots,x_n)$ with $\sum m_ix_i=0$:
$$\2={1\over M}\pmatrix{M-m_1& -m_1 &-m_1 &\dots& -m_1\cr -m_2&M-m_2&-m_2&\dots &-m_2\cr -m_3 & -m_3 &M-m_3&\dots& -m_3\cr \dots&\dots&\dots&\dots&\dots
\cr
-m_n&-m_n&-m_n&\dots&M-m_n},\qquad  M=\sum_{i=1}^n m_i.$$

{\bf 2.5. Definition.} The {\it shifted Wintner-Conley matrix} of a central configuration with positive masses is the matrix $\Zs=\Z-\lambda\2$.

{\bf 2.6. Properties.} The basic properties of $\Zs$ are the following. We have $0=(1,\dots,1)\,\Zs$. The configuration $q$ is a central configuration with multiplier $\lambda$ if and only if $q\Zs=0$. The $i$-th column $$(u\Zs)_i=\sum_{j,j\neq i}m_j(S_{ij}-{\lambda\over M})(u_i-u_j),$$ where $u=(u_1,\dots,u_n)\in(\R^p)^n$. Let $\mu$ be the mass matrix $$\mu=\pmatrix{m_1& 0  &\dots& 0\cr 0&m_2&\dots &0\cr  \dots&\dots&\dots&\dots
\cr
0&0&\dots&m_n}.$$ Then $\Z\mu$, $\2\mu$ and $\Zs\mu$ are symmetric. Let  $v=(v_1,\dots,v_n)\in(\R^p)^n$. We get 
$$\sum_{i,j,i<j}m_im_j(S_{ij}-{\lambda\over M})\langle u_i-u_j,v_i-v_j\rangle=\trace({}^t\!v u\Zs\mu ).$$

{\bf Proof of Proposition 1.6.} According to $(2.2)$, a vector $u=(u_1,\dots,u_n)$ belongs to the kernel of the Hessian quadratic form if and only if $g=(g_1,\dots,g_n)=0$. If $u$ is vertical, the second term of $g$ vanishes. The first term $u\Zs$ should vanish. Each row of this matrix should vanish. Assume for example a configuration $(2.4)$ in $\O xyz$ which is 2-dimensional, conventionally in the $\O xy$ plane, satisfying consequently $z_1=\cdots=z_n=0$. Let
$$u=(u_1,u_2,\dots, u_n)=\pmatrix{\xi_1&\xi_2&\cdots&\xi_n\cr
\eta_1&\eta_2&\cdots&\eta_n\cr \zeta_1&\zeta_2&\cdots&\zeta_n}.$$
The first two rows are zero, and $(\zeta_1,\dots,\zeta_n)$ is a disposition in the left kernel of $\Zs$. If the planar configuration was in $\R^4$ instead of $\R^3$, the fourth coordinate would produce another disposition in the same kernel. QED

{\bf Proof of Proposition 1.5.} Assume that a vector $u=(u_1,\dots,u_n)$ is in the kernel of the Hessian quadratic form. It satisfies $g=(g_1,\dots,g_n)=0$. Let us take the same 3-dimensional example as in the previous proof.  The projection of $u$ on the vertical direction $\O z$ is
$$\hat u=\pmatrix{0&0&\cdots&0\cr
0&0&\cdots&0\cr \zeta_1&\zeta_2&\cdots&\zeta_n}.$$
If we replace $u$ by $\hat u$ in $(2.2)$, we find again $g_i=0$, since the $x$ and $y$ coordinates of $g_i$ are sum of two zero terms, and the $z$ coordinates are unchanged.
QED

A vector $u$ in the kernel of the Hessian quadratic form gives $\hat u$ in the same kernel, and $\zeta=(\zeta_1,\dots,\zeta_n)$ in the left kernel of $\Zs$. If $\hat u$ is nontrivial in the sense of \S 1, then a $\zeta$ is nontrivial in the sense that it is not a linear combination of the known vectors in the left kernel of $\Zs$.
We may restate this fact as follows.

{\bf 2.7. Proposition.} A central configuration $q$ satisfies $\rank \Zs+\rank q\leq n-1$. It has a nontrivial vertical degeneracy if and only if $\rank \Zs+\rank q\leq n-2$.

{\bf Proof.} In the left kernel of $\Zs$ we find the vector $(1,\dots,1)$ and, since $q\Zs=0$, each row of the matrix $q$, which gives the first inequality. According to $(2.2)$, $-\Zs\mu$ is the vertical Hessian quadratic form. The vector $(1,\dots,1)$ generates an infinitesimal translation of the configuration. The rows of $q$ generate infinitesimal rotations. If there are other vectors in the left kernel of $\Zs$, there are other degeneracies of the vertical Hessian quadratic form: these are nontrivial vertical degeneracies. QED

\bigskip

\centerline{\bf 3. Changes of dimension. Two proofs of Proposition 1.4.}

\bigskip

{\bf Short proof of Proposition 1.4.} We consider a sequence $q^\kappa$, $\kappa\in\N$, of configurations of dimension $>d$, which converges to a $d$-dimensional configuration $q^\infty$, and a sequence of mass vectors $m^\kappa$, which converges to a mass vector $m^\infty$. These configurations are central. In particular, they do not have collisions and the mass vectors have positive entries. According to 2.6, $q^\kappa\Zs^\kappa=0$ and $q^\infty\Zs^\infty=0$. The rank is lower semi-continuous. The rank of $\Zs$ cannot increase in the limit. According to Proposition 2.7, $\Zs^\infty$ has a nontrivial kernel, i.e., the central configuration has a vertical degeneracy. QED

{\bf Constructive argument.} The previous argument does not tell what is the nontrivial kernel. Clearly, it is associated to a deformation of the configuration which changes the dimension.

Here is an argument which gives a vector in the kernel. Again, we consider a sequence $q^\kappa$, $\kappa\in\N$, of configurations in a space of dimension $>d$, which converges to a $d$-dimensional configuration $q^\infty$. We renumber the bodies such that the $d+1$ first bodies of $q^\infty$ form a nonflat simplex (of dimension $d$). Then by a small rotation and a small translation of each configuration $q^\kappa$ we obtain another sequence $\hat q^\kappa$ which also converges to $q^\infty$ and has the following property: (i) for any $\kappa\in\N$, for any $i=1,\dots, d+1$, the first $i$ bodies of $\hat q^\kappa$ generate the same affine subspace as the first $i$ bodies of $q^\infty$. So, whatever $\kappa\in\N$, $\hat q_1^\kappa=q^\infty_1$, $\hat q_2^\kappa$ is on the line $q^\infty_1q^\infty_2$, $\hat q^\kappa_3$ is on the plane $q^\infty_1q^\infty_2q^\infty_3$, etc. The bodies $\hat q_1^\kappa,\dots, \hat q_{d+1}^\kappa$ are in the ``horizontal'' space. All their ``vertical'' coordinates are zero. The vertical coordinates of the other bodies converge to zero.

Assume now that the configurations $q^\kappa$ and $q^\infty$ are all central. There is now a sequence of mass vectors $m^\kappa$. Our hypothesis is that for any $i$, $i=1,\dots,n$, $m_i^\kappa\to m_i^\infty>0$.  Consequently also  the  multipliers $\lambda^\kappa\to \lambda^\infty>0$, and the total mass $M^\kappa\to M^\infty>0$. We rescale to whole sequence in order to get $\lambda^\infty=M^\infty$. By slightly rescaling again each $q^\kappa$ we get this second property: (ii) for any $\kappa\in\N$, $\lambda^\kappa=M^\kappa$.

Applying both the small rotation and the change of scale, we obtain from the sequence $q^\kappa$ another sequence with the properties (i) and (ii). In what follows we will assume, without loss of generality, that the sequence $q^\kappa$ has the above properties (i) and (ii).

The hypothesis of Proposition 1.4 is: for any $\kappa$, $\dim q^\kappa>d$. This hypothesis implies that after renumbering the $n-d-1$ last bodies, at least one of the vertical coordinates of $q_n^\kappa$  takes infinitely many values when $\kappa\to+\infty$. Let us call $t$ such a coordinate. It will be the coordinate $x_l$ of the following proposition, in the case $x_l(t)=t$.

{\bf 3.1. Proposition (\citealt{hampton}).} Let $\T=(\C\setminus\{0\})^r$. Suppose that a system of $s$ polynomial equations $f_i(x)=0$ defines an infinite variety $V\subset \T$. Then there is a nonzero rational vector $\alpha= (\alpha_1,\dots,\alpha_r)$, a point $a = (a_1,\dots,a_r)\in \T$, and Puiseux series $x_j(t) = a_j t^{\alpha_j} +\cdots$, $j = 1,\dots, r$, convergent in some punctured neighborhood $U$ of $t=0$, such that $f_i(x_1(t),\dots,x_r(t))=0$ in $U$, $i =1,\dots,s$. Moreover, if the projection from $V$ onto the $x_l$-axis is dominant, there exists such a series solution with $x_l(t)=t$ and another with $x_l(t)=t^{-1}$.

The usual existence results of Puiseux series are too weak for our purpose, while this general proposition is exactly what we need. We apply it to the system $q\Zs=0$ which characterizes the central configurations and which is also: $\sum_{j\neq i}m_j(S_{ij}-1)(q_i-q_j)=0$, for any $i=1,\dots,n$. Recall that $S_{ij}=\|q_i-q_j\|^{-3}$. The system will be polynomial if the variables are the vectors $q_1,\dots, q_n$, the numbers $m_1,\dots,m_n$ and the numbers $S_{ij}$, $1\leq i<j\leq n$. We add to the system defining the central configuration the polynomial constraints $S_{ij}^2\|q_i-q_j\|^6=1$.

We consider a neighborhood of the limiting $d$-dimensional central configuration $q^\infty$. We consider as the new variables, suitable with Proposition 3.1, the increments $q_i-q^\infty_i$, $m_i-m^\infty_i$, $S_{ij}-S_{ij}^\infty$ of the old variables. We restrict the increments of the $q_i$'s, $i=1,\dots,d+1$, in agreement with property (i). The system is again polynomial in these new variables.  In order to agree with the hypothesis $V\subset \T$, we remove from the list of variables any scalar increment which is identically zero. Then, we can apply 3.1 to the new list of variables.

{\bf  3.2. Proposition.} If an $n$-body central configuration $q^\infty$ of dimension $d$ is the limit of a sequence of $n$-body central configurations of dimension $>d$, then, after renumbering the bodies, there is a  Taylor series in $\varepsilon\in\C$ converging near $\varepsilon=0$ and parametrizing central configurations by expressing the configuration matrix as $q=q^\infty+\varepsilon q^{[1]}+\varepsilon^2q^{[2]}+\cdots$ and the mass vector as $m=m^\infty+\varepsilon m^{[1]}+\varepsilon^2m^{[2]}+\cdots$. Moreover, the $q_i$'s, $i=1,\dots,d+1$, generate a $d$-dimensional affine space which is independent of the value of $\varepsilon$, which we call the horizontal space. One of the vertical coordinates of $q_n$ is $\varepsilon^k$, for some positive integer number $k$.

Note that Proposition 3.1 states the existence of a Puiseux series while Proposition 3.2 states the existence of a Taylor series.  Indeed the variable of 3.2 is $\varepsilon=t^{1/k}$ where $t$ is given in 3.1. Moreover, there are no negative powers in the series since we are dealing with increments which are small as $\varepsilon\to 0$.

{\bf Constructive proof of Proposition 1.4.} We expand the equation of central configurations until the first order $i$ in $\varepsilon$ such that  $q^{[i]}$ is not horizontal (the case $i=1$ will reprove 1.5). We write
$$0=q\Zs=(q^\infty+\varepsilon q^{[1]}+\varepsilon^2q^{[2]}+\cdots)(\Zs^\infty+\varepsilon\Zs^{[1]}+\varepsilon^2\Zs^{[2]}+\cdots).$$
Call $z=(z_1,\dots,z_n)\in\R^n$ the list of vertical coordinates of $q=(q_1,\dots,q_n)$ that is nonzero at order $i$: $z^\infty=z^{[1]}=\cdots=z^{[i-1]}=0$, $z^{[i]}\neq 0$. The above equation is true after replacing the matrix $q$ by its row $z$. Consequently $z^{[i]}\Zs^\infty=0$.
Note that Proposition 3.1 does not discuss the existence of a real branch, so $z^{[i]}$ may not be real. But then its real part and its imaginary part are both in the left kernel of the real matrix $\Zs^\infty$. So, we have a vertical degeneracy. Property (i) as restated in Proposition 3.2 implies that $z_1=\cdots=z_{d+1}$ at any order in $\varepsilon$. If $z^{[i]}$ was a linear combination of the lists of horizontal coordinates of $q^\infty$, the vectors $q_j^\infty-q_1^\infty$, $j=2,\dots,d+1$, would be linearly dependent. But we assumed the contrary. The vertical degeneracy is nontrivial. QED

\bigskip

\centerline{\bf 4. Context of Theorem 1.2}

\nobreak
\bigskip

We first recall two inequalities which are used to determine a sign in Dziobek's formulas. We consider the matrix $\Z=(\Z_{ij})_{1\leq i,j\leq n}$ of $(2.3)$ and the matrix $\Z-\lambda\2=\Zs=(\Zs_{ij})_{1\leq i,j\leq n}$.

{\bf 4.1. Proposition (\citealt{albouy3}).} Consider a central configuration with positive masses. Then for any $i$, $j$, $i\neq j$, $\Zs_{ij}+\Zs_{ji}\leq \Zs_{ii}+\Zs_{jj}$, with equality if and only if all the other bodies are equidistant from bodies $q_i$ and $q_j$.

We will not reproduce the proof. Here are alternative expressions. The nonnegative quantity $ \Zs_{ii}+\Zs_{jj}-\Zs_{ij}-\Zs_{ji}=
\Z_{ii}+\Z_{jj}-\Z_{ij}-\Z_{ji}-2\lambda$. It is also, with the notation $\Sigma_i$ of $(2.3)$, $$\Sigma_i+\Sigma_j+(m_i+m_j)S_{ij}-2\lambda=2(m_i+m_j)S_{ij}+\sum_{k\neq i,j}m_k(S_{ik}+S_{jk})-2\lambda,$$ or finally, with the notation $\Ss_{ij}=S_{ij}-\lambda/M$, $2(m_i+m_j)\Ss_{ij}+\sum_{k\neq i,j}m_k(\Ss_{ik}+\Ss_{jk})$. Moeckel deduced his Theorem 1.3 from the following proposition.

{\bf 4.2. Proposition (\citealt{moeckel}).} For any central configuration with positive masses, $\trace \Zs\geq 0$. The equality only occurs for the equilateral simplex, for which $\Zs=0$.

{\bf Proof.} This is a corollary of Proposition 4.1. We know that $\sum_i \Zs_{ij}=0$. Consequently $-\trace \Zs$ is the sum of all the nondiagonal entries of $\Zs$. Summing up the inequality of 4.1 for all $i$, $j$, $i<j$, we find $-\trace \Zs\leq (n-1)\trace \Zs$. QED

{\bf 4.3. Previous works.}  \citet{wintner} introduces the shifted matrix $
\Gamma=\lambda\1-\Z$ at $\S356$. He expresses the equation of central configurations as we did. He writes ``Actually, $r$ is always precisely the multiplicity of the root $0$ of the characteristic equation of $\Gamma$.'' There is a misprint since Wintner defined $r$ as the rank of $\Gamma$. \citet{albouy3} claims that Wintner was stating the impossibility of a ``vertical degeneracy''. This was indeed assuming that the misprinted $r$ was the rank of the other matrix, which represents the configuration. But more probably the misprinted $r$ is simply $n-r$ and Wintner's statement is trivial. This paragraph is somewhat neglected by Wintner, which is consistent with his conclusion: ``Nevertheless, the characterization  of central configurations in terms of the matrix $\Gamma$ is often unmanageable, and will not be used in what follows.'' However, in \citet{pacella}, \citet{albouychenciner} and in what follows, several results are deduced from this characterization.

This short paragraph by Wintner may be inspired by the long article \citet{meyer}, where the eigenvectors of the same matrix are discussed. At \S 24, on page 152, Meyer claims to prove that for a collinear central configuration of $n$ bodies, the matrix $\Z$, which he writes on page 133, has $n-2$ eigenvalues strictly greater than $\lambda$. The vertical degeneracy of a collinear central configuration is impossible. These two results are true, as later proved by Conley and published in \citet{pacella}. Meyer's argument is the last paragraph of his \S 24. He translates the lower rank possibility into an equivalent statement about the kernel, which he expresses through the Krediet--Laura--Andoyer equations\footnote{\citet{krediet} seems to have been the first author to publish the so-called Andoyer equations for an arbitrary number of bodies in the plane. See his equation (8).}. But he does not explain why this should be ``impossible''. In \citet{palmore}, the same claim is made, also with unclear arguments ``by comparing the terms''. 

Previously \citet{saari} had sketched an argument that would discard the vertical degeneracy in any dimension. \citet{moeckel} showed a counter example, a configuration made of two regular polygons of 473 bodies each. \citet{moeckelsimo} further describe this configuration. Note that bifurcations of 5 bodies configurations are common in the plane (see \citealt{meyerschmidt} or \citealt{xia}) and in 3 dimensions (see \citealt{acll}, \citealt{smpv}).

{\bf Dziobek configurations and flat Dziobek configurations.} A {\it Dziobek configuration} is, according to the most usual definition, an $n$-body central configuration of dimension exactly $n-2$. According to Proposition 2.7, the shifted Wintner-Conley matrix $\Zs$ is of rank one (since $\rank\Zs=0$ gives dimension $n-1$, see \S 6).

{\bf 4.4. Definition.} A {\it flat Dziobek configuration} with positive masses is an $n$-body central configuration of dimension $\leq n-3$ such that $\rank\Zs=1$.

We may restate Theorem 1.2 as: {\it there are no flat Dziobek configurations with $n=5$}. For $n\geq 6$, we do not know if there are flat Dziobek configurations.  They would be central configurations with a vertical degeneracy. 

{\bf Remark.} \citet{albouy3} extends the usual definition of a Dziobek configuration: They are defined by the condition $\rank \Zs=1$. With this definition, flat Dziobek configurations would be Dziobek configurations. See also the remark of \S 7 in this reference, concerning the hypothesis of positive masses, and note an inconsistency concerning this definition: the proof of Proposition 12 indeed assumes the usual definition.

{\bf 4.5. Proposition.} A configuration with positive masses is a central configuration satisfying $\rank \Zs=1$ if and only if there is a $\lambda\in\R$, a nonzero $(\Delta_1,\dots,\Delta_n)\in\R^n$ such that $\sum\Delta_i=0$, $\sum\Delta_iq_i=0$, and, for all $i$ and $j$, $i<j$, $m_im_j(S_{ij}-\lambda/M)=-\Delta_i\Delta_j$.

{\bf Proof.} As the matrix $\Zs\mu$ is symmetric of rank one, it is of the form $\pm(\Delta_i\Delta_j)_{i,j}$.  As $(1,\dots,1)\Zs=0$, $\sum_i\Delta_i=0$. The condition for a central configuration is $q\Zs=0$, which is equivalent to $\sum_i\Delta_iq_i=0$. According to Proposition 4.2, the $\pm$ sign is $+$. QED

{\bf 4.6. Proposition (\citealt{dziobek, schmidt}).} An $n$-body configuration of dimension exactly $n-2$, with positive masses, is central if and only if there is a $\lambda\in\R$, a nonzero $(\Delta_1,\dots,\Delta_n)$ such that $\sum\Delta_i=0$, $\sum\Delta_iq_i=0$, and, for all $i$ and $j$, $m_im_j(S_{ij}-\lambda/M)=-\Delta_i\Delta_j$.

{\bf Proof.} We have $q$ and $(1,\dots,1)$ in the left kernel, so $\Zs$ is of rank one. We conclude by Proposition 4.5. QED

{\bf Remark.} If the dimension of a configuration is exactly $n-2$, the first two conditions $\sum\Delta_i=0$, $\sum\Delta_iq_i=0$ determine uniquely, up to a factor, a nonzero $(\Delta_1,\dots,\Delta_n)$. We call the $\Delta_i$'s the {\it homogeneous barycentric coordinates} of the configuration.

{\bf Remark.} This proof of Dziobek's main formula is really short. The idea is from \citet{albouy2}. Its simplicity contradicts Wintner's opinion (see 4.3). The original proof by Dziobek relies on an identity involving Cayley's determinant, which is often considered as difficult (\citealt{saari2}). Dziobek indeed made a sign mistake (\citealt{corsroberts}). \citet{brehm}, \citet{meyer}, \citet{wintner} and \citet{meyerschmidt} praised Dziobek's work, but they did not write his identity nor his main formula. \citet{schmidt} writes them and extends them to the case of a 3-dimensional 5-body configuration. \citet{moeckel2} gives a simple proof of the extended Dziobek identity together with a very elementary way to prove Dziobek's main formula without Cayley determinant (the same as advertized later in \citealt{albouy3}).

{\bf Remark.} If $\rank \Zs=1$ and if $S_{ij}=\lambda/M$, then $\Delta_i=0$ or $\Delta_j=0$ and consequently at least $n-1$ of the mutual distances are $(\lambda/M)^{-1/3}$. The same relation $S_{ij}=\lambda/M$ is still remarkable if  $\rank \Zs>1$, but the conclusions are not as strong. \citet{gidea}, \citet{chh} independently discovered a one-parameter family of 5-body planar central configurations with $q_2-q_1=q_3-q_2$ and $S_{24}=S_{25}=\lambda/M$.

{\bf 4.7. Proposition (\citealt{albouy3}).} In a  central configuration $q_1,\dots,q_n$ with $\rank \Zs=1$ satisfying $\Delta_n=0$, the bodies $q_1,\dots,q_{n-1}$ are equidistant from $q_n$ and they form a central configuration with $\rank \Zs=1$.

{\bf Proof.} We simply make $\Delta_n=0$ in the equations of Proposition 4.5. We get the same equations with body $n$ ignored and, for all $i\leq n-1$, $S_{in}=\lambda/M$. QED

The normalized multiplier $\lambda/M$ of the $n-1$-body central configuration should be the normalized multiplier of the $n$-body central configuration since in both cases, if $1\leq i <j\leq n-1$, $m_im_j(S_{ij}-\lambda/M)=-\Delta_i\Delta_i$. Note that $M$ is different in both cases. Such configurations do exist for 5 bodies in dimension 3. They were introduced by Erich Ludwig Brehm. This reciprocal statement may be extracted from Brehm's formulas:

{\bf 4.8. Proposition (\citealt{brehm}, p.\ 28).} Given any 4-body cocircular central configuration $(q_1,\dots,q_4)$ with positive masses $(m_1,\dots,m_4)$ and a mass $m_5>0$, there exists, up to reflection, a unique nonplanar 5-body central configuration $(q_1,\dots,q_5)$ with same $q_1,\dots,q_4$ and with masses $(m_1,\dots,m_5)$. In this configuration, $r_{15}=r_{25}=r_{35}=r_{45}$. This common distance does not depend on $m_5$. The height $h$ of the resulting pyramid satisfies
$r_{15}\sqrt2/2<h<r_{15}\sqrt3/2$.

{\bf Proof.} Let us first see the uniqueness. A 3-dimensional 5-body central configuration satisfies the conditions of Proposition 4.6. The first four bodies generate an affine plane, so they have homogeneous barycentric coordinates, i.e., there is a $(\Delta_1,\dots,\Delta_4)\neq 0$ such that $\sum_{i=1}^4\Delta_i=0$ and $\sum_{i=1}^4\Delta_iq_i=0$. Joining $\Delta_5=0$, we get the homogeneous barycentric coordinates $(\Delta_1,\dots,\Delta_5)$ of the 5-body configuration.  As $\Delta_5=0$, for $i<5$, $S_{i5}=r_{i5}^{-3}=\lambda/M$. So the pyramid is unique. About the existence, note that if the distance $r_{15}=(\lambda/M)^{-1/3}$ was too small, the pyramid would be impossible. So the existence is a consequence of a lower bound on $r_{15}$ or on the height $h$. A cocircular configuration is convex. Assume the cyclic order $1234$. As explained by Dziobek, $\Delta_1,\Delta_3<0<\Delta_2,\Delta_4$. Then $m_1m_3(r_{13}^{-3}-r_{15}^{-3})=-\Delta_1\Delta_3<0$ and $m_1m_2(r_{12}^{-3}-r_{15}^{-3})=-\Delta_1\Delta_2>0$, so $r_{12}<r_{15}<r_{13}$. The length of a side of the convex quadrilateral is smaller than $r_{15}$, which is smaller than the length of a diagonal. If the radius of the circle is $r$, the greatest side should be greater than $\sqrt2 r$, and the diagonals smaller than $2r$. So, $\sqrt2 r<r_{15}<2r$, or $2(r_{15}^2-h^2)<r_{15}^2<4(r_{15}^2-h^2)$, which gives Brehm's inequalities. In another way, $r<h<\sqrt 3 r$. QED

{\bf 4.9. Proposition (\citealt{asv}).} A flat Dziobek 5-body central configuration with positive masses and with $\Delta_5=0$ is impossible.

{\bf Proof.} Proposition 4.5 gives us the $\Delta_i$'s. Then we can repeat the argument in the proof of 4.8. The planar configuration would have $h=0$, which is excluded by the inequalities. This is a proof. Let us give a second proof. Proposition 4.7 shows that the configuration has four bodies on a circle and the fifth body at the center. Theorem 7 of \citet{asv} states that for such configuration $\rank \Zs=2$.  QED

\bigskip

\centerline{\bf 5. Proof of Theorem 1.2.}

\bigskip

{\bf Idea of the proof.} Assuming the existence of a flat Dziobek configuration we will show a contradiction between the equations $\sum\Delta_i=0$, $\sum \Delta_iq_i=0$, which constrain $(\Delta_1,\dots,\Delta_5)$, and Dziobek's equations $S_{ij}=z-\delta_i\delta_j$, $1\leq i<j\leq 5$, where $\delta_i=\Delta_i/m_i$.   In other words, we will find contradictions between the ``barycentric point of view'' (we call the $\Delta_i$'s the homogeneous barycentric coordinates of the $q_i$'s) and the ``Distance-Dziobek point of view''.

A well-known example of such a contradiction proves the statement: {\it A 2-dimensional 4-body central configuration cannot have three bodies on a line.} By Proposition 4.6 there would be a $(\Delta_1,\dots,\Delta_4)$ with 
$\sum\Delta_i=0$, $\sum \Delta_iq_i=0$, $S_{ij}=z-\delta_i\delta_j$. If the three points $q_1$, $q_2$, $q_3$ are on a line, $\Delta_4=0$ according to the first two equations. But $\delta_4=0$ and $S_{ij}=z-\delta_i\delta_j$ implies that $q_1$, $q_2$, $q_3$ are on a circle (this is indeed a case of Proposition 4.7). This is a contradiction: since $q_1$, $q_2$, $q_3$ are  distinct points, they cannot be on a circle and on a line. The same kind of contradiction allows to prove the symmetry of the equal mass 4-body central configurations (see \citealt{albouy1}).

\bigskip

{\bf Barycentric point of view.} We shall describe the possible $(\Delta_1,\dots,\Delta_5)$ for a planar 5-body configuration. They form a plane $\Pi\subset\R^5$ defined by the three independent linear equations  $\sum\Delta_i=0$, $\sum \Delta_ix_i=0$, $\sum \Delta_iy_i=0$.
We shall describe all the possible planes $\Pi$ and all the possible combinations of  signs of the $\Delta_i$'s on each plane.

If we had $\Delta_1=0$ on all a plane $\Pi$, there would be a plane of homogeneous barycentric coordinates $(\Delta_2,\dots,\Delta_5)$ of bodies 2 to 5. These bodies would consequently be collinear. This is impossible in a 2-dimensional 5-body central configuration, according to the perpendicular bisector theorem. So $\Delta_1=0$ defines a vectorial line in $\Pi$. The plane $\Pi$ is divided into sectors by the conditions $\Delta_i=0$, $i=1,\dots,5$. The question is to determine in which cyclic order these conditions happen. Two lines $\Delta_i=0$ and $\Delta_j=0$ may coincide, as we will see in Table B.
A coincidence of three lines corresponds to a collision $q_i=q_j$. But collisions are excluded in our problem.

\bigskip
\centerline{\includegraphics[width=70mm]{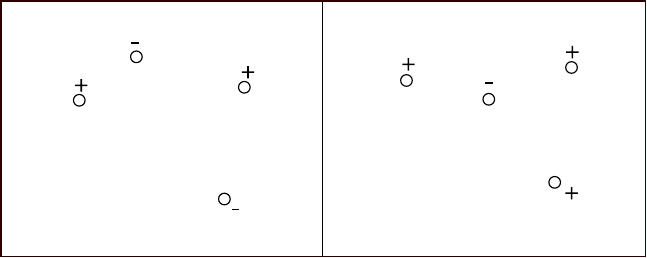}}
\centerline{Figure 1. Barycentric signs}
\bigskip

When for example $\Delta_1=0$, we observe the configuration formed by bodies $2345$. As $(\Delta_2,\dots,\Delta_5)$ are homogeneous barycentric coordinates of this configuration, the $\Delta_i$'s should have consistent signs, illustrated in Figure 1. If the quadrilateral is not convex, the exterior signs are the same and they differ from the interior sign: the interior body has positive barycentric coordinates in the barycentric frame of the exterior bodies. If the quadrilateral is convex, the signs should alternate cyclically (compare \citealt{saari2}, \S 3.3).

$$\matrix{1&2&3&4&5\cr 0&+&-&+&-\cr +&+&-&+&-\cr +&0&-&+&-\cr +&-&-&+&-\cr +&-&0&+&-\cr +&-&+&+&-\cr +&-&+&0&-\cr +&-&+&-&-\cr +&-&+&-&0\cr +&-&+&-&+\cr 0&-&+&-&+\cr}\quad\left|\quad\matrix{1&2&3&4&5\cr 0&+&-&+&-\cr +&+&-&+&-\cr +&0&-&+&-\cr +&-&-&+&-\cr +&-&0&+&-\cr +&-&+&+&-\cr +&-&+&+&0\cr  +&-&+&+&+\cr +&-&+&0&+\cr +&-&+&-&+\cr 0&-&+&-&+\cr}\quad\right|\quad
\matrix{1&2&3&4&5\cr 0&+&-&+&-\cr +&+&-&+&-\cr +&+&0&+&-\cr +&+&+&+&-\cr +&0&+&+&-\cr +&-&+&+&-\cr +&-&+&+&0\cr  +&-&+&+&+\cr +&-&+&0&+\cr +&-&+&-&+\cr 0&-&+&-&+\cr}$$
\centerline{\includegraphics[width=111mm]{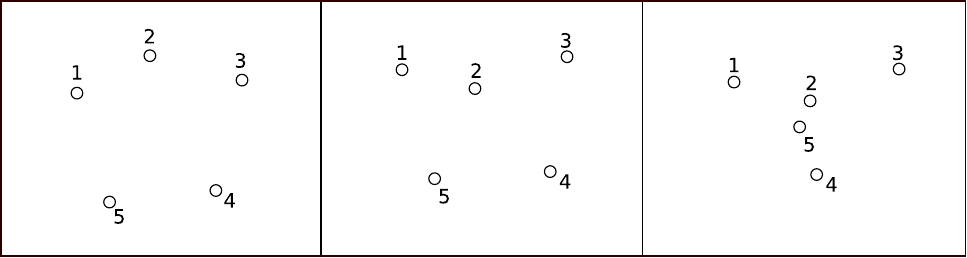}}
\centerline{Table A}
  
\bigskip

$$\matrix{1&2&3&4&5\cr 0&+&-&+&-\cr +&+&-&+&-\cr +&0&-&+&-\cr +&-&-&+&-\cr +&-&0&+&-\cr +&-&+&+&-\cr +&-&+&0&0\cr  +&-&+&-&+\cr 0&-&+&-&+}
\quad\left|\quad
\matrix{1&2&3&4&5\cr 0&+&-&+&-\cr +&+&-&+&-\cr +&0&0&+&-\cr +&-&+&+&-\cr +&-&+&+&0\cr   +&-&+&+&+\cr +&-&+&0&+\cr +&-&+&-&+\cr 0&-&+&-&+}\right.$$\centerline{\includegraphics[width=70mm]{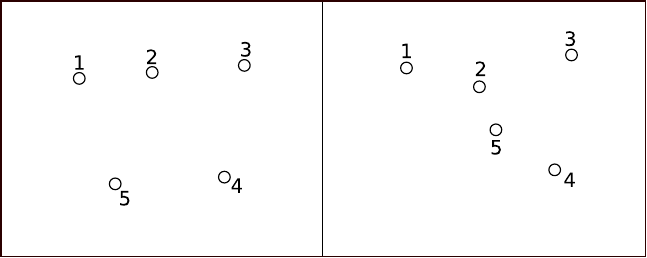}}
\centerline{Table B}

\bigskip

Suppose first that the 5-body configuration is convex, in cyclic order 12345. If $\Delta_1=0$ then $(\Delta_2,\dots,\Delta_5)$ are homogeneous barycentric coordinates
of bodies 2 to 5, which form a convex configuration in the cyclic order 2345. Consequently their signs are $+-+-$ or $-+-+$. These two choices are equivalent. We write these signs as the first row of Table A1. The following reasoning gives the other rows of Table A1. The second row is a small perturbation of the condition $\Delta_1=0$, which gives $\Delta_1>0$ and does not change the other signs. The case $\Delta_1<0$ would give the same table in reverse order. In the third row the only choice is a 0 at body 2, since only this choice gives correct signs $+-+-$ for the remaining bodies, which also form a convex quadrilateral. We check at each step that this reasoning gives a unique choice for the next vanishing $\Delta_i$. The last row is the first row with all the signs changed: by continuing we would follow the same sequence again with all the signs changed.

In Table B1 we assume that bodies 123 are on a line in this order, 1345 being convex in this cyclic order. In Table A2, body 2 passes slightly inside. As $\Delta_5=0$, we find the signs $+-++$ of a concave configuration. In Table B2, we push body 5 on the side 14.  We see that the cases with three collinear bodies are obtained from a generic case by suppression of an intermediate step. {\it The allowed nongeneric cases do not produce new combinations of five signs compared to the generic cases.} The last type of generic case is described as the configuration in Table A3, a triangle with 134 outside, and 25 inside, such that 2345 is convex in this cyclic order. The other two quadrilaterals which include 2 and 5 are concave.

{\bf Remark.} Esther Klein remarked in 1933 that by extracting a chosen point of a 5-point planar configuration one can leave a convex quadrilateral. She assumed that 3 points are not on a same line.  The question was first extended in \citet{erdos}.

\bigskip

 {\bf Distance-Dziobek point of view.} We have obtained all the combinations of signs of the $\Delta_i$'s. We will now see their incompatibilities with Dziobek's equations $S_{ij}=z-\delta_i\delta_j$. The method is inspired from \citet{fernandes}. We recall our notation: $r_{ij}=\|q_i-q_j\|$ and $S_{ij}=r_{ij}^{-3}$, $\delta_i=\Delta_i/m_i$.

{\bf 5.1. Proposition.} A convex planar configuration of 4 points $(q_1,q_2,q_3,q_4)$ in this cyclic order, such that there exists $z>0$ and $(\delta_1,\delta_2,\delta_3,\delta_4)$ with signs $++--$, with $S_{ij}=z-\delta_i\delta_j$, is impossible.

{\bf Proof.} The $r_{ij}$ are ordered as the $\delta_i\delta_j$. There would be two large distances $r_{12}$ and $r_{34}$, corresponding to opposite sides. Let  the smallest of both be $r_{12}$. According to the four small distances $r_{13}$, $r_{23}$, $r_{14}$, $r_{24}$, and to the cyclic order, bodies 3 and 4 should be both in an ogive (see Figure 2) delimited by the segment 12 and by two circles, centered respectively in bodies 1 and 2, passing respectively through 2 and 1. But the greatest distance in the ogive is $r_{12}$. The largest side 34 does not fit in the ogive. Contradiction. QED

\bigskip

\centerline{\includegraphics[width=40mm]{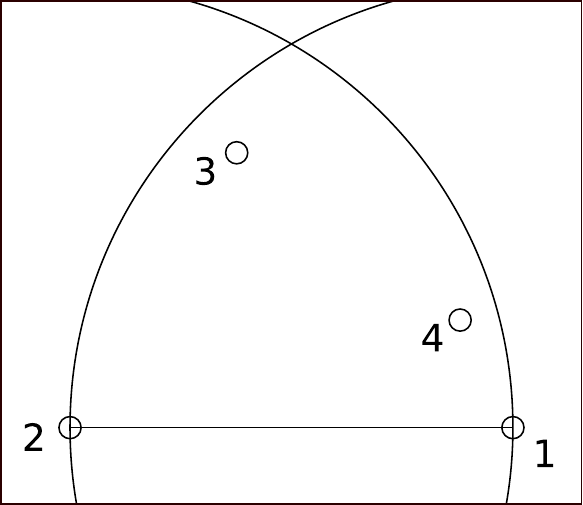}$\qquad\qquad$\includegraphics[width=40mm]{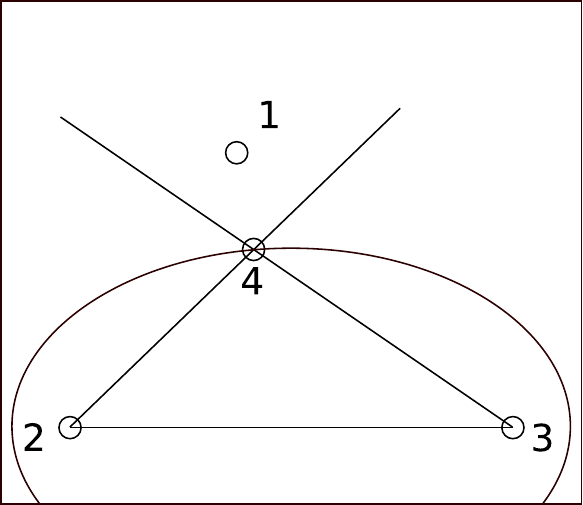}}
\centerline{Figure 2. Proofs of 5.1 and 5.2}

\bigskip

{\bf 5.2. Proposition.} A concave planar configuration of 4 points $(q_1,q_2,q_3,q_4)$, the fourth being inside the triangle of the first three, such that there exists $z>0$ and $(\delta_1,\delta_2,\delta_3,\delta_4)$ with 
signs $-+++$, with $S_{ij}=z-\delta_i\delta_j$, is impossible.

{\bf Proof.} Due to the concavity hypothesis, $q_1$ is ``behind'' (see Figure 2) the vertex $q_4$ of the triangle $q_2q_3q_4$. But $r_{12}$ and $r_{13}$ are ``small'' distances, since $\delta_1\delta_2$ and $\delta_1\delta_3$ are negative, while $r_{24}$ and $r_{34}$ are ``great'' distances, since $\delta_2\delta_4$ and $\delta_3\delta_4$ are positive. This is a contradiction. More precisely, we may observe that $r_{24}+r_{34}<r_{12}+r_{13}$ since the ellipse with foci $q_2$ and $q_3$, passing through $q_4$, is bisecting the angle 243.  QED

{\bf 5.3. Proposition.} If a convex planar configuration of 4 points $(q_1,q_2,q_3,q_4)$ is such that there exists $z>0$ and $(\delta_1,\delta_2,\delta_3,\delta_4)$ such that $0<\delta_4\leq\delta_3\leq\delta_2\leq \delta_1$ and that $S_{ij}=z-\delta_i\delta_j$, then the cyclic orders may be 1432 or 1324. In the first case, the angle $132$ is acute and we have the estimate $2r_{34}<r_{12}$. In the second case, the angle $132$ is obtuse and $\sqrt{3}r_{34}\leq r_{12}$.

{\bf Proof.} We choose coordinates $\O xy$ with $q_2=(-1,0)$ and $q_1=(1,0)$. The $r_{ij}$ are ordered as the $\delta_i\delta_j$: $$r_{34}\leq r_{24}\leq (r_{14}\hbox{ or }r_{23})\leq r_{13}\leq r_{12}.$$
The point $q_3=(x_3,y_3)$ is in the half-ogive of Figure 3, with $x_3\leq 0$ and $y_3\geq 0$. We draw the circle of center $q_2$ passing through $q_3$ and the level curve of $$f(x,y)=\bigl((x+1)^2+y^2\bigr)^{-3/2}-\bigl((x-1)^2+y^2\bigr)^{-3/2}$$ passing through $q_3$. We have $f(x_3,y_3)\geq 0$. For any positive value of $f$ the level curve is a closed curve on the left half-plane. The only intersection of the level curve and the circle in the upper semi-plane is $q_3$. Here, the value of $f$ on a point moving on the circle counterclockwise from the $\O x$ axis to the $\O x$ axis is increasing.

 \bigskip
\centerline{\includegraphics[width=50mm]{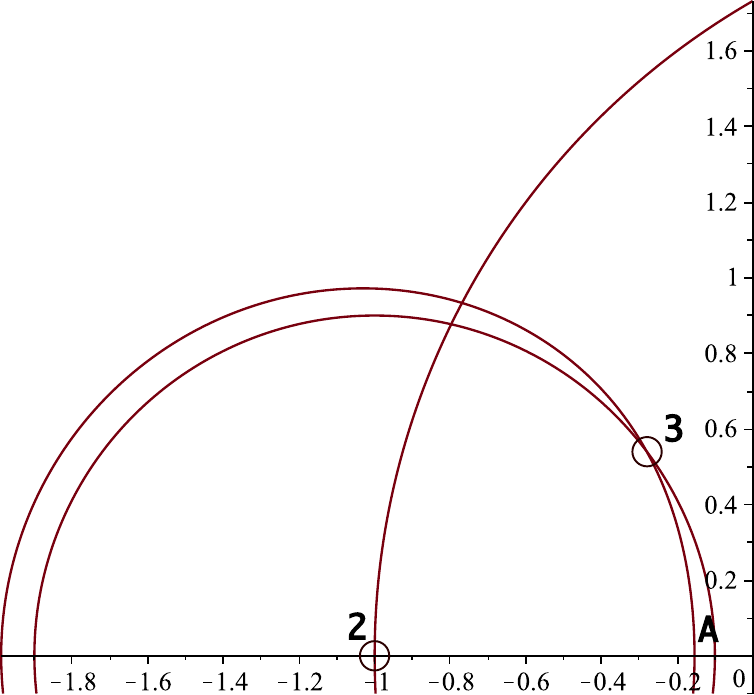}$\qquad\qquad$\includegraphics[width=50mm]{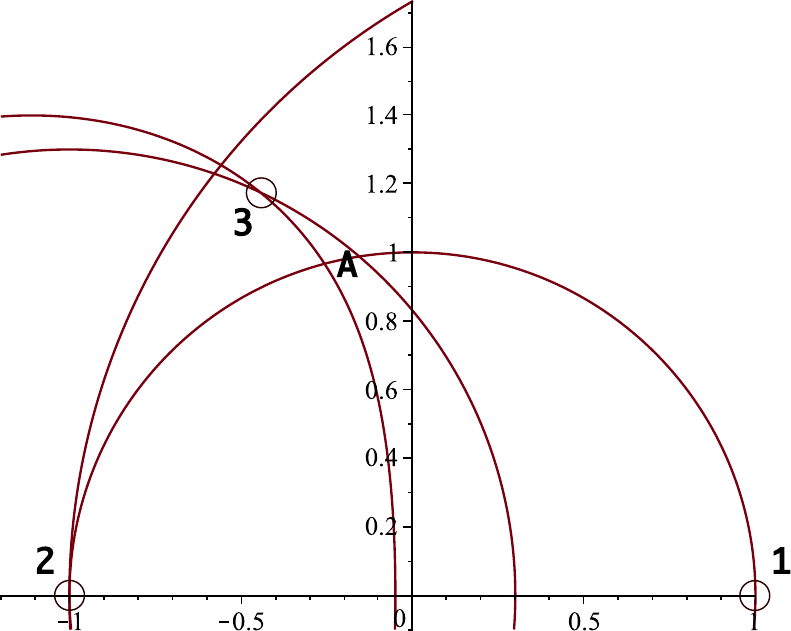}}
\nobreak

\centerline{Figure 3. Proof of 5.3, first case}

\bigskip

{\bf First case. The point $q_4$ is in the upper half plane.} If $\delta_1=\delta_2$, then $x_3=x_4=0$ and the configuration is not convex. So $\delta_2<\delta_1$. The point $q_4$ should be in the same half-ogive as $q_3$, inside the circle (since $r_{24}<r_{23}$) and outside the level curve since
$${S_{24}-S_{14}\over S_{23}-S_{13}}={\delta_4\over \delta_3}\leq 1,$$
which means that $f(x_4,y_4)\leq f(x_3,y_3)$.
It belongs to the region A (see Figure 3, left). But the convexity condition requires in particular  $q_4$ above the triangle $q_1q_3q_2$. The region A is inside this triangle, except if $q_3$ is above the unit circle of the plane. In this case, the angle $q_2q_3q_1$ is acute, and the segment $q_3q_1$ crosses the disk of center $q_2$ passing through $q_3$ (see Figure 3, right). 
To get a simple bound on $r_{34}$, we observe that $q_4$ is located between the circle of center $q_2$ passing through $q_3$, the segment from $q_3$ to $q_1$ and the $\O y$ axis. Within this domain, due to the negative slope, the farthest point from $q_3=(x_3,y_3)$ is the intersection of the segment with $\O y$, if the domain is cut by $\O y$. If the domain is not cut by $\O y$, this distance will still be an upper bound of $r_{34}$. This bound is $r_{13}x_3/(x_3-1)$. It is maximal at $q_3=q_2$ where it is $r_{12}/2$.

{\bf Second case. The point $q_4$ is in the lower half plane.} The domain for the point $q_4$ is restricted by the perpendicular bisector of $(q_2,q_3)$, by the $\O x$ and $\O y$ axes and by the circle of center $q_2$ passing through $q_3$.
Due to the perpendicular bisector, $q_3$ should be within the unit disk
(see Figure 4). The angle $132$ is obtuse. The perpendicular bisector crosses the $\O y$ axis at $y=(x_3^2+y_3^2-1)/(2y_3)<0$. As the distance $r_{34}$ is increased when going farther along the perpendicular bisector, the point $q_4=(0,y)$ maximizes the distance to $q_3$. The distance $r_{34}$ is again maximized if $x_3=0$ and $y_3=\sqrt{3}/3$, in which case $x_4=0$, $y_4=y=-y_3$ gives the upper bound $2\sqrt{3}/3$ for the distance $r_{34}$. If $y_3<\sqrt{3}/3$, the constraint by the circle give $y_4>-y_3$ and a shorter distance $r_{34}$. If $y_3>\sqrt{3}/3$, the constraint by the perpendicular bisector gives $y_4>(y_3^2-1)/(2y_3)$ and a shorter distance $r_{34}$. QED

 \bigskip
\centerline{\includegraphics[width=70mm]{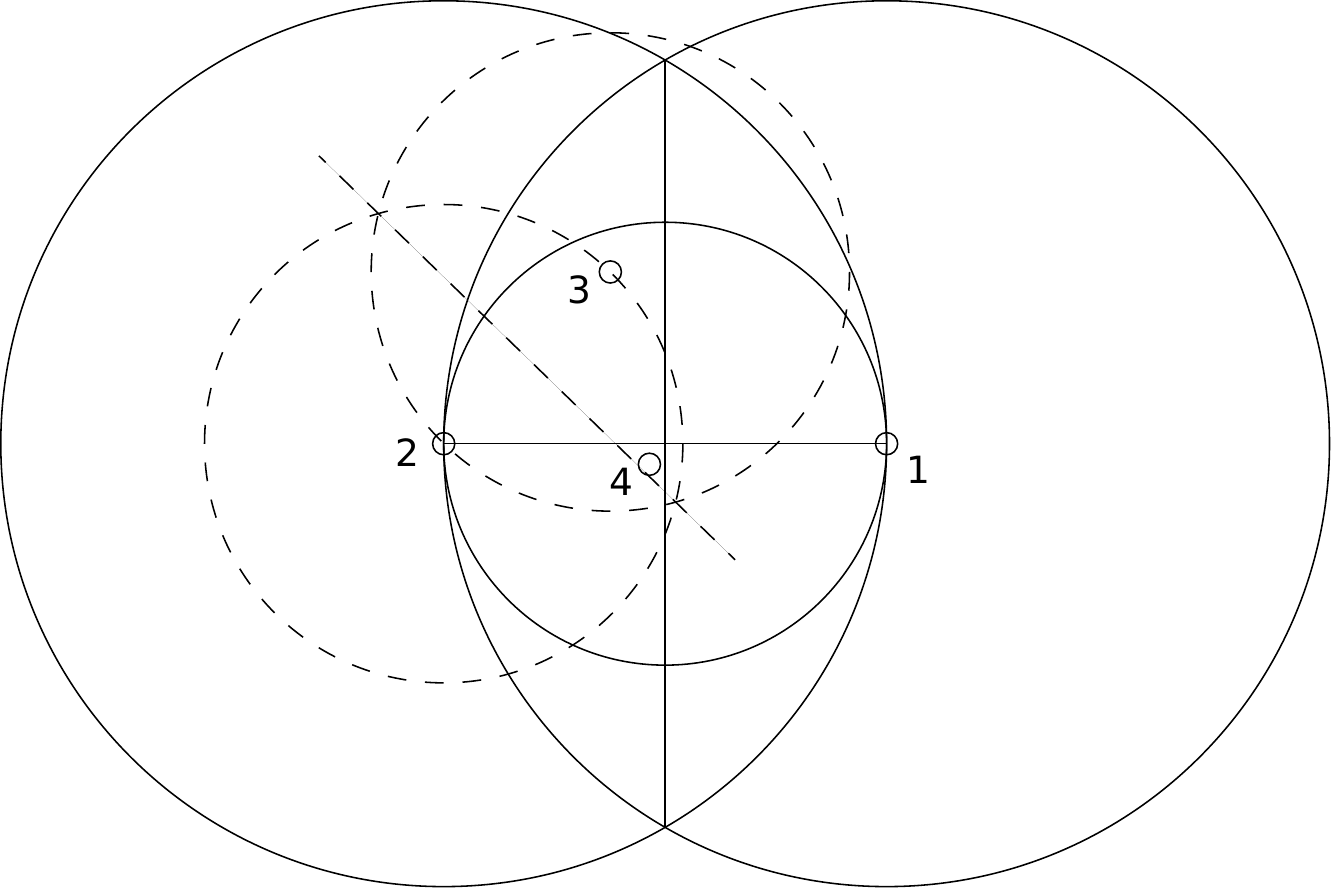}}

\centerline{Figure 4. Proof of 5.3, second case}
\bigskip

{\bf 5.4. Proposition.} A planar configuration of 5 points $(q_1,q_2,q_3,q_4,q_5)$ such that the first four points form a convex quadrilateral with the fifth point inside, and such that there exists $z>0$ and $(\delta_1,\delta_2,\delta_3,\delta_4,\delta_5)$ with signs $++++-$, with $S_{ij}=z-\delta_i\delta_j$, is impossible.

 \bigskip
\centerline{\includegraphics[width=70mm]{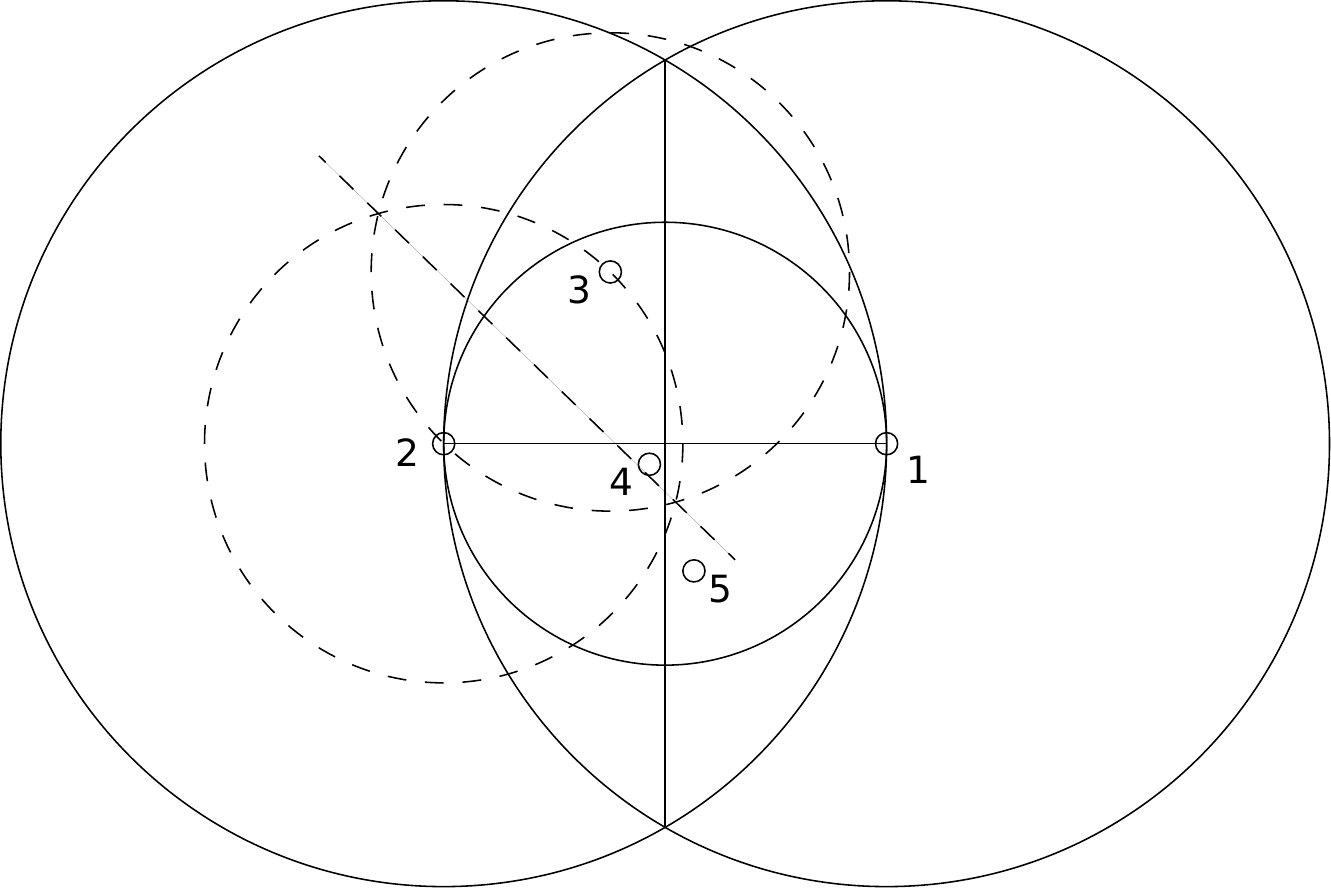}}

\centerline{Figure 5. Proof of 5.4}

\bigskip

{\bf Proof.} We may renumber the points in such a way that $\delta_5<0<\delta_4\leq\delta_3\leq\delta_2\leq \delta_1$. The order of the distances satisfies
$$(r_{15}\hbox{ or }{r_{12}\over 2})\leq r_{25}\leq r_{35}\leq r_{45}<r_{34}\leq r_{24}\leq (r_{14}\hbox{ or }r_{23})\leq r_{13}\leq r_{12}.$$ The inequality
$r_{12}/2\leq r_{25}$ comes from the triangular inequality $r_{12}\leq r_{15}+r_{25}$. Proposition 5.3 applies to the first four points. Since $r_{12}/2< r_{34}$, the first case is excluded. In the second case, we should have $x_5>0$
and $q_5$ should be below the perpendicular bisector of $q_2q_3$ (see Figure 5). In particular, $y_5<y_4$. So, $q_5$ is not in the interior of the convex quadrilateral $q_1q_3q_2q_4$. QED

{\bf Impossibility of the 5-body flat Dziobek configurations.} We assume that there are 5 noncollinear, non coinciding, positions $q_1,\dots,q_5$ in the plane, a positive number $z$ and 5 nonzero numbers $\delta_1,\dots,\delta_5$, such that $S_{ij}=z-\delta_i\delta_j$, where $S_{ij}=\|q_i-q_j\|^{-3}$.

Furthermore, we assume that 5 homogeneous barycentric coordinates $\Delta_i$, $i=1,\dots,5$ of the 5 positions $q_i$ are such that $\Delta_i/\delta_i>0$.

These two conditions are satisfied for a flat Dziobek configuration with positive masses $m_1,\dots, m_5$. In this case we have $\Delta_i/\delta_i=m_i$ and $z=\lambda/M$ where $\lambda$ is the multiplier of the central configuration and $M=m_1+\cdots+m_5$. A zero $\delta_i$ (or $\Delta_i$) is impossible according to Proposition 4.9. We will examine the rows without zeros in Table A. We do not need to examine Table B or other non-generic cases since the rows without zero are the same, and the impossibilities of Propositions 5.1, 5.2 and 5.4 are easily extended to the boundary defined by the collinearity of 3 bodies.

Table A1, second row, is impossible, since if we ignore body 4, the remaining quadrilateral satisfies the hypothesis of Proposition 5.1. The other rows of this table are similar. In the fourth row, we ignore body 5. In the sixth row, we ignore body 1 (note that this ignored $+$ sign is equidistant from the zeros of the same column.) Etc.

Table A2, second row, is impossible, since if we ignore body 5, the quadrilateral satisfies the hypothesis of Proposition 5.2. The fourth row is similar, ignoring body 4. The sixth row is impossible according to Proposition 5.1, ignoring body 1. The eighth row is excluded by Proposition 5.4. The tenth row is excluded by 5.1, ignoring body 3.

Table A3, second row, is impossible according to 5.2, ignoring body 5. The fourth row is excluded by 5.2 ignoring body 4. The sixth row, by 5.1 ignoring body 1. The eighth row, by 5.2 ignoring body 3. The tenth row, by 5.2 ignoring body 2. Theorem 1.2 is proved.

\bigskip

\centerline{\bf 6. Proof of Theorem 1.1.}

\bigskip

Let us begin with the changes of dimension which are impossible whatever the number $n$ of bodies.

The maximal dimension for an $n$-body configuration is $n-1$. In this dimension the equation for central configurations $q\Zs=0$ implies $\Zs=0$ and $S_{ij}=\lambda/M$ for any $i,j$, $1\leq i<j\leq n$ (compare \citealt{saari}). The configuration is a regular simplex of sides $r_{ij}=(\lambda/M)^{-1/3}$. Clearly, a central configuration of dimension $\leq n-2$ cannot be a limit of regular simplexes of dimension $n-1$.

The minimal dimension of a configuration without collision is one. As mentioned in \S 4.3, Conley proved that a nontrivial vertical degeneracy of a collinear central configuration is impossible. According to Proposition 1.4, a limit of central configurations of dimension $d>1$ cannot be of dimension one. A stronger result is known, which delimitates, independently of the masses, a neighborhood of the collinear central configurations where no central configuration of higher dimension can be found: According to \citet{moeckel}, if a noncollinear central configuration is given, one cannot find a direction $L$ such that for all $i,j, 1\leq i<j\leq n$, the angle $(q_iq_j,L)\leq \pi/4$.

These results are sufficient to exclude the changes of dimension if $n=4$. If $n=5$, there only remains to prove that a planar central configuration cannot be a limit of spatial central configurations. This is proved by Theorem 1.2 and Proposition 1.4.

\bigskip

{\bf Acknowledgements.} This work has benefited from various discussions in January 2023, at the seminar organized at UFPE, Recife, by Eduardo S.G. Leandro. We thank Alan A.\ Santos, Marcelo P.\ Santos and the reviewer for their useful comments.


{\bf Statements and Declarations.} We wish to thank the Brazilian-French Network in Mathematics, the projects FAPEMIG APQ 03149-18 and CNPq 306568/2021-7 for giving us opportunities to meet. 

\end{document}